\documentclass[superscriptaddress,reprint,amssymb, nobibnotes, aps, pra]{revtex4-2}
\usepackage{graphicx}
\usepackage{amsmath}
\usepackage{float}
\usepackage{braket}
\usepackage{bibentry}

\usepackage{color,soul}
\usepackage{appendix}

\begin{document}

\title{Stochastic Analysis of Laser Phase Noise in Photonics-Assisted Radio Frequency Beamforming}

\author{Niloy Ghosh}
\affiliation{Smart Frontend Group, Netherlands Institute for Radio Astronomy, Dwingeloo, The Netherlands}
\affiliation{Photonics Wireless Convergence Lab, Electro-Optical Communication Systems, Eindhoven University of Technology, The Netherlands}

\author{David Prinsloo}
\affiliation{Smart Frontend Group, Netherlands Institute for Radio Astronomy, Dwingeloo, The Netherlands}

\author{Eduward Tangdiongga}
\affiliation{Photonics Wireless Convergence Lab, Electro-Optical Communication Systems, Eindhoven University of Technology, The Netherlands}

\begin{abstract}
This paper develops a stochastic model for quantifying the impact of laser phase noise on photonics-assisted radio frequency (RF) beamforming. Unlike conventional deterministic models, the presented theoretical framework captures the statistical randomness associated with the RF array factor of beamforming networks driven by finite-linewidth lasers. Fourier-domain analysis reveals that laser phase noise reshapes the spatial coherence function of such beamforming networks, inducing coherent RF aperture shrinkage, beam-pattern distortion, and directivity degradation. The analysis highlights that beamforming network architectures employing nanosecond-scale optical differential delays can exhibit substantial performance degradation even when driven by commercially available narrow-linewidth (kHz-range) lasers. 
Furthermore, it is demonstrated that average RF power gain gets downscaled and fluctuations dominate in large-aperture beamforming networks due to phase noise, imposing an upper limit on the maximum achievable RF antenna array-size. The reported results highlight the quantitative performance and scalability limits imposed by laser phase noise in photonics-assisted RF beamforming, offering key design guidelines for implementing large-scale beamforming networks in next-generation wireless communications, sensing, radar, and radio astronomy applications.
\end{abstract}

\maketitle

\section{Introduction}

Beamforming is a key enabling technology for high-capacity wireless communication, sensing, radar, and radio astronomy \cite{10144712,zhao2026simulation,dongre2025robust}. Although digital beamforming offers high flexibility, the requirement for a dedicated electronic RF-chain, analog-to-digital converter, and digital-to-analog converter per antenna channel significantly increases the hardware complexity, power consumption, and cost \cite{dutta2019case}. On the other hand, analog beamforming alleviates these challenges, but suffers from limited bandwidth, beam squint, and electromagnetic interference due to the use of electronic phase-shifters \cite{marpaung2019integrated}. These limitations have motivated growing interest in photonics-assisted RF beamforming, which offers broadband beam steering with minimized beam squint, low loss, and immunity to electromagnetic interference.

Optical true time delay (OTTD) is the core building block of photonics-assisted RF beamforming networks. Significant progress has been made on the implementation of OTTD lines using a wide range of photonic technologies, including fiber-based, wavelength-division multiplexing-based, integrated photonics-based, and microcomb-based approaches \cite{duan2023optical,jung2007optical,martinez2024ultrabroadband,yu2014multi,xue2018microcomb,zhang2021large,sun2022broadband,gao2016simultaneously,porzi2021fast,serafino2020high}. These advances have further enabled integrated and reconfigurable multi-beamforming based on Butler and Blass matrix architectures for phased-array systems \cite{lu2021integrated,zhou2024butler,reza2021design,loukisa2026blass}.

Despite significant progress in the development of photonics-assisted RF beamforming networks, the impact of laser phase noise on such systems has received very little attention. Conventionally, ideal phase coherence is assumed between the constituent OTTD channels of beamforming networks driven by finite-linewidth lasers. However, this assumption is fundamentally inaccurate as tunable differential delays are intentionally incorporated between the channels to achieve beam steering. This can decorrelate the phase fluctuations across the OTTD channels, causing significant performance degradation. Although self-homodyne detection can suppress the effect of laser phase noise in coherent optical communication links by canceling common phase fluctuations \cite{morsy2018dsp,ghosh2019complete,mecozzi2016kramers}, this approach typically relies on having negligible path-length mismatch between the optical channels, and hence cannot be directly applied to beamforming networks. A stochastic framework quantifying the impact of laser phase noise on RF beamforming is still missing.

Understanding and analyzing the influence of laser phase noise becomes even more critical as beamforming networks continue to scale towards larger antenna arrays and longer optical delay ranges for future wireless communication, sensing, and radio astronomy applications. Such systems may experience substantial phase decorrelations across the constituent OTTD channels even when driven by narrow-linewidth lasers, potentially limiting their performance. Quantifying these limits is therefore essential for selecting laser specifications, determining the maximum achievable array-size, and practical design trade-offs for large-scale photonics-assisted RF beamforming networks.

In this paper, we develop a rigorous stochastic framework for analyzing the impact of finite laser linewidth on photonics-assisted RF beamforming networks. Unlike conventional deterministic formulations, the proposed framework models the RF array factor of a beamforming network as a stochastic process governed by random laser phase noise. Based on this formulation, it is shown that laser phase noise reshapes the spatial coherence function of the beamforming network. Fourier-domain analysis interprets that reshaping of spatial coherence function in turn shrinks the effective RF aperture of the beamforming network, causes RF beam-pattern distortion, gain reduction, and directivity collapse. The analysis is further extended to characterize the limits on system metrics like coherent array gain, beam fluctuation, and beamforming fidelity, imposed by laser phase noise, thereby establishing fundamental system performance and scalability limits. Moreover, this paper shares key quantitative guidelines for designing large-scale photonics-assisted RF beamforming systems operating under finite laser linewidth. 

The remainder of this paper is organized as follows. Section II develops the stochastic model of OTTD-based RF beamforming in the presence of laser phase noise in the system. Section III presents the Fourier-domain analysis of spatial coherence function to study the impact of laser phase noise on the coherent aperture of the beamforming network and on the resultant RF beam-pattern. Section IV analyzes the impact of differential delay and array-dimension on different beamforming performance metrics, and discusses the resulting design implications accounting for finite laser linewidth.

\section{Stochastic modeling of Photonics-assisted RF Beamforming under laser phase noise}

\begin{figure}[ht]
    \centering
    \includegraphics[width=\linewidth]{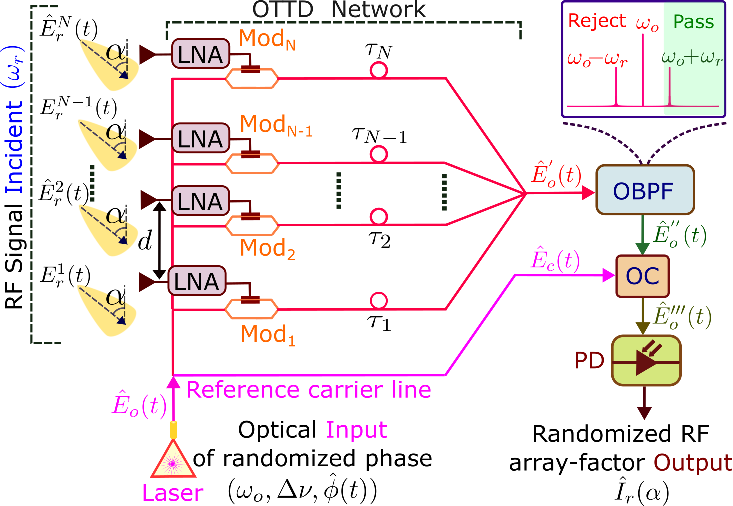}
    \caption{Architecture of optical true time delay based RF beamforming network operating in the receiving mode. LNA: Low Noise Amplifier, OBPF: Optical Bandpass Filter, OC: Optical Combiner, PD: Photodetector.}
    \label{Figure1}
\end{figure}

Fig.~\ref{Figure1} illustrates the schematic of an OTTD based RF beamforming network operating in receiving mode. The network comprises of an $N$-element antenna array with inter-element spacing $d$. A plane RF wave arriving from direction $\alpha$ is received by the array, producing at the $n^{\mathrm{th}}$ antenna the following RF electric field,
\begin{equation}
E^{n}_r(t)= |E_r| \cos{\big(\omega_r t +(n-1) k_r d \cos{\alpha}\big)}
\label{Eq1}
\end{equation} 
where $|E_r|$ is the electric field strength, $\omega_r$ is the angular frequency, and $k_r$ is the propagation constant of the received RF signal. 
Each antenna output is first amplified by a low noise amplifier (LNA), which subsequently drives an optical intensity modulator. All modulators are driven by a common continuous-wave laser source of frequency $\omega_o$ and linewidth $\Delta \nu$. In this work, the laser phase noise $\hat{\phi}(t)$ is modeled as a zero-mean Wiener process, which accurately describes the phase evolution of practical semiconductor lasers \cite{agrawal2012fiber}. Table~\ref{TableCh2} summarizes all the variables used in this paper and the parameter they represent. 

\begin{figure}[ht]
    \centering
\includegraphics[width=\linewidth]{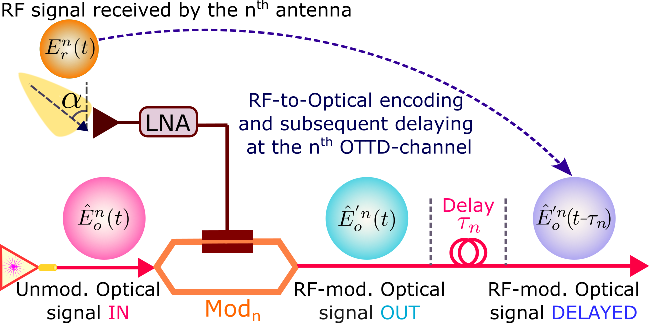}
    \caption{Optical amplitude modulation with the received RF signal, followed by adding delay of $\tau_n$ at the $n^{\mathrm{th}}$ OTTD channel of the beamforming network.}
    \label{Figure1_2}
\end{figure}

The optical modulators in the system are indexed as [Mod$_1$, Mod$_{2}$....Mod$_{N-1} $, Mod$_{N}$]. The optical electric field at the output of the laser $\hat{E}_o(t)$ is equally distributed into ($N$+1) branches. The first $N$ optical branches drive the antenna-coupled modulators, whereas the remaining branch carries the optical reference carrier for coherent detection. We denote the optical signal at the input of the $n^{\mathrm{th}}$ modulator i.e. Mod$_n$ as $\hat{E}^{n}_{o}(t)$=$|E_o| e^{j(\omega_o t + \hat{\phi}(t))}$. Mod$_n$ performs linear amplitude modulation of the optical signal $\hat{E}^{n}_{o}(t)$ with the RF signal $E^{n}_r(t)$ received by the $n^{\mathrm{th}}$ antenna, as illustrated in Fig.~\ref{Figure1_2}. The resultant RF-amplitude modulated optical signal $\hat{{E}}^{'n}_{o}(t)$ at the output of Mod$_{n}$ can be expressed as,
\begin{equation}
\hat{{E}}^{'n}_o(t) = \big(1+ E^{n}_r(t) \big)  |E_o| e^{j(\omega_o t + \hat{\phi}(t))}
\label{Eq3}
\end{equation}
The modulated optical signals at the output of each modulator are subsequently routed through parallel OTTD channels. The delays incorporated by each OTTD channel is indexed as [$\tau_1$, $\tau_{2}$....$\tau_{N-1}$, $\tau_{N}$]. Here, $\tau_n$=$n\tau$ refers to the delay of the $n^{\mathrm{th}}$ OTTD channel, where $\tau$ is the inter-channel differential delay. The time-delayed modulated optical signal $\hat{{E}}^{'n}_o(t-\tau_{n})$ at the $n^{\mathrm{th}}$ OTTD channel can be expressed as,
\begin{equation}
\hat{{E}}^{'n}_o(t-\tau_{n}) = \big(1+ E^{n}_r(t-\tau_n) \big)  |E_o| e^{j(\omega_o (t-\tau_{n}) + \hat{\phi}(t-\tau_{n}))}
\label{Eq4}
\end{equation}

The optical signal $\hat{E}'_o(t)$ correspondingly obtained from the coherent addition of all OTTD channels is,

\begin{equation}
    \hat{E}'_o(t) = \sum_{n=1}^{N} \big(1+ E^{n}_r(t-\tau_n) \big)  |E_o| e^{j(\omega_o (t-\tau_{n}) + \hat{\phi}(t-\tau_{n}))}
\label{Eq5_2}
\end{equation}
\begin{figure*}[ht]
    \centering
    \includegraphics[width=\linewidth]{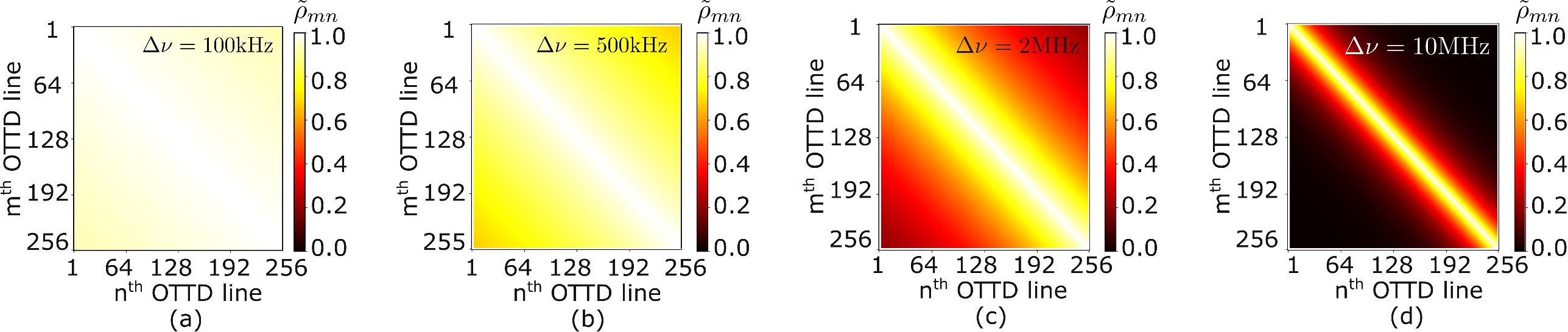}
    \caption{Heatmap plots of the correlation matrix $[\tilde{\rho}]$ of an OTTD-based RF beamforming network of array-dimension $N$=256 and differential delay  $\tau$=1ns, driven by a laser of linewidth (a) $\Delta \nu$=100kHz, (b) $\Delta \nu$=500kHz, (c) $\Delta \nu$=2MHz, and (d) $\Delta \nu$=10MHz.}
    \label{Figure4_2}
\end{figure*}
\begin{figure*}[ht]
    \centering
    \includegraphics[width=\linewidth]{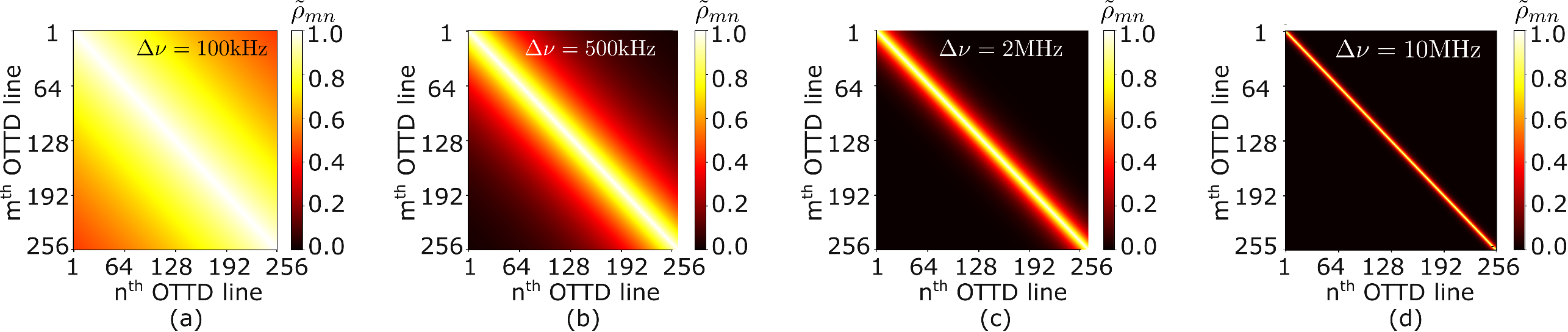}
    \caption{Heatmap plots of the correlation matrix $[\tilde{\rho}]$ of an OTTD-based RF beamforming network of array-dimension $N$=256 and differential delay  $\tau$=10ns, driven by a laser of linewidth (a) $\Delta \nu$=100kHz, (b) $\Delta \nu$=500kHz, (c) $\Delta \nu$=2MHz, and (d) $\Delta \nu$=10MHz.}
    \label{Figure4_3}
\end{figure*}
 \noindent The optical signal $\hat{E}'_o(t)$ is then subjected to bandpass filtering in order to retain the spectral components centered at the upper sideband frequency $(\omega_o+\omega_r)$. The optical output of the bandpass filter $\hat{E}^{''}_o(t)$ can be expressed as follows,
\begin{equation}
\hat{E}^{''}_o(t)= \sum_{n=1}^{N} \kappa e^{j((\omega_o + \omega_r) (t-\tau_n) +(n-1) \gamma_{\alpha} + \hat{\phi}(t-\tau_n))}
\label{Eq6}
\end{equation}
where the constant $\kappa$ in the above equation is ($|E_o| |E_r|/2$), and  $\gamma_{\alpha}$ is the RF phase-lag between two consecutive OTTD channels which can be expressed as,
\begin{equation}
\gamma_{\alpha} = \omega_r \tau + k_r d \cos{\alpha}
\label{Eq7_2}
\end{equation} 

 The filtered optical signal $\hat{E}^{''}_o(t)$ is further subjected to carrier-reinsertion by combining it with the unmodulated signal $\hat{E}^{c}_n(t)$=$|E_o| e^{j(\omega_o t + \hat{\phi}(t-\tau_c))}$ from the reference carrier line. Here, $\tau_c$ corresponds to the time-delay incorporated by the reference carrier line. Finally, the optical signal $\hat{E}^{'''}_o(t)$=$\hat{E}^{''}_o(t)+\hat{E}_c(t)$ post carrier-reinsertion is subjected to photodetection. The normalized RF-photocurrent $\hat{I}_r(\alpha) \sim\hat{E}^{*'''}_o(t)\cdot\hat{E}^{'''}_o(t)$ at the photodetector output can be derived to be,
\begin{equation}
\hat{{I}}_r(\alpha) = \sum_{n=1}^{N}   e^{j (\hat{\phi}(t-\tau_n) - \hat{\phi}(t-\tau_c))} \cdot  e^{j (n-1) \gamma_{\alpha} } 
\label{Eq7_1}
\end{equation}

Equation~(\ref{Eq7_1}) constitutes the central result of the stochastic system model. In the absence of laser phase noise, the stochastic exponential term $e^{j (\hat{\phi}(t-\tau_n) - \hat{\phi}(t-\tau_c))}$ in (\ref{Eq7_1}) reduces to unity, and the detected photocurrent corresponds to the conventional deterministic RF array factor \cite{balanis2016antenna}. Under finite laser linewidth, however, the differential delays introduced in the OTTD channels of the beamforming network cause the laser phase to decorrelate between channels. Consequently, the RF array factor becomes a stochastic quantity whose statistics is jointly governed by the laser linewidth and the OTTD channel delays. This stochastic formulation forms the core foundation of the remainder of the paper. Leveraging the derived formulation, we will investigate the impact of laser phase noise on RF beamforming in the upcoming sections.  

\begin{table}[h!]
\centering
\caption{Summary of variables and the corresponding parameter they respresent.}
 \begin{tabular}{||c| c |c||} 
 \hline
\textbf{Sl.}  & \textbf{Variable} & \textbf{Parameter}\\ [0.5ex] 
 \hline
 1 & $\hat{\phi}(t)$ & Laser phase noise random process\\
  \hline
2 & $\Delta \nu$ & Spectral linewidth of laser\\
 \hline
3 & $\tau_n$ & Delay of $n^{\mathrm{th}}$ OTTD channel \\
 \hline
4 & $\tau$ & Differential delay b/n neighbouring channels\\
 \hline
5 & $\alpha$ & RF reception angle\\
\hline
6 & $\alpha_o$ & Direction of RF beam maxima\\
\hline
7 & $\gamma_{\alpha}$ & Inter-channel RF phase lag\\
\hline
8 & $\hat{I}_r(\alpha)$ & RF field array factor random process\\
\hline
9 & $\hat{P}_r(\alpha)$ & RF power array factor random process\\
\hline
10 & $\braket{\hat{{P}}_r(\alpha)}$ & Mean RF power array factor\\
\hline
11 & $\tilde{\rho}_{mn}$ & Average correlation b/n $m^{\mathrm{th}}$ and $n^{\mathrm{th}}$ channels\\
\hline
12 & $\tilde{\rho}$ & Average correlation b/n neighbouring channels\\
\hline
13 & $\tilde{C}_x$ & Spatial coherence function\\
\hline
14 & $x$ & Spatial index\\
\hline
15 & $\braket{\hat{G}_r}$ & Mean RF power gain\\
\hline
16 & ${\tilde{g}_r}$ & Gain-scaling factor\\
\hline
17 & $G_c$ & Average coherent array gain\\
\hline
18 & $\Delta G_v$ & Gain variance\\
\hline
19 & $\tilde{F}$ & Beamforming fidelity\\
\hline
\end{tabular}
\label{TableCh2}
\end{table}

%derive the resulting inter-channel coherence function and present a spatial Fourier-domain analysis, demonstrating how laser-induced coherence decay translates into effective RF aperture reduction and beam-pattern degradation. %Building on this framework, Section IV 

\section{Influence of laser phase noise on RF beamforming}

\subsection{Inter-channel coherence decay}

In this subsection, we characterize laser phase noise induced coherence decay across the OTTD channels of the beamforming network. To proceed, we first derive the statistical properties of the RF power array factor. From the stochastic RF field array factor presented in (\ref{Eq7_1}), the corresponding RF power array factor $\hat{P}_r(\alpha)$=$|\hat{I}_r(\alpha)|^{2}$ can be derived to be,
\begin{equation}
\hat{{P}}_r(\alpha) = \sum_{m=1}^{N}  \sum_{n=1}^{N}   e^{j (\hat{\phi}(t-\tau_m) - \hat{\phi}(t-\tau_n))} \cdot  e^{j (m-n) \gamma_{\alpha} } 
\label{Eq8}
\end{equation}
The stochastic term $e^{j (\hat{\phi}(t-\tau_m) - \hat{\phi}(t-\tau_n))}$ in ~(\ref{Eq8}) is associated with laser phase noise induced phase-decorrelation between the $m^{\mathrm{th}}$ and $n^{\mathrm{th}}$ OTTD channels, for all $(m,n) \in [1, N]$. The mean RF power array factor $\braket{\hat{{P}}_r(\alpha)}$ can be derived to be {(see Appendix-A)},
\begin{equation}
\braket{\hat{{P}}_r(\alpha)} =  \sum_{m=1}^{N} \sum_{n=1}^{N}  \tilde{\rho}_{mn}  e^{j (m-n) \gamma_{\alpha} }
\label{Eq10}
\end{equation}
where the parameter $\tilde{\rho}_{mn}$ quantifies the average correlation between the $m^{\mathrm{th}}$ and $n^{\mathrm{th}}$ OTTD channels, and is related to the delay difference $\tau |m-n|$ between the corresponding channels and laser linewidth $\Delta \nu$ as,
\begin{equation}
\tilde{\rho}_{mn}=e^{-\pi \Delta \nu \tau |m-n|}
\label{Eq10_2}
\end{equation}
The $m^{\mathrm{th}}$ and $n^{\mathrm{th}}$ OTTD channels are: (a) Completely correlated if $\tilde{\rho}_{mn}$=1,  (b) Partially uncorrelated if $0<\tilde{\rho}_{mn}<1$,  and (c) Completely uncorrelated if $\tilde{\rho}_{mn}$=0.
Equation(\ref{Eq10_2}) reveals that the linewidth-delay product $\Delta \nu \tau$ emerges as the governing parameter controlling spatial coherence across the all channels of the beamforming network. Channels differing by larger delays experience progressively weaker mutual correlation, whereas neighboring channels remain relatively correlated.

The correlation matrix $ [{\tilde{\rho}}]$ represented below offers an intuitive visualization of linewidth-delay product governed coherence loss between the different OTTD channels in a beamforming network,
\begin{equation}
    [{\tilde{\rho}}] =  \begin{bmatrix}
          1 &  e^{-\pi \Delta \nu \tau }  &  \textbf{.} & \textbf{.} &  e^{-(N-1)\pi \Delta \nu \tau }\\
         e^{-\pi \Delta \nu \tau } & 1  & \textbf{.} & \textbf{.}  & e^{-(N-2)\pi \Delta \nu \tau }\\
         e^{-2\pi \Delta \nu \tau } & e^{-\pi \Delta \nu \tau } & \textbf{.} & \textbf{.} &   e^{-(N-3)\pi \Delta \nu \tau }\\
        e^{-3\pi \Delta \nu \tau } & e^{-2\pi \Delta \nu \tau }  & \textbf{.}  & \textbf{.} &   e^{-(N-4)\pi \Delta \nu \tau }\\
         \textbf{.} & \textbf{.} & \textbf{.} & \textbf{.}  & \textbf{.} \\
         \textbf{.} &  \textbf{.} & \textbf{.}  & \textbf{.} & \textbf{.}\\
          e^{-(N-1)\pi \Delta \nu \tau } & e^{-(N-2)\pi \Delta \nu \tau } &  \textbf{.} & \textbf{.} &   1\\
        \end{bmatrix} 
\label{Eq16}
\end{equation}

In the ideal coherent limit $(\Delta \nu \tau \rightarrow 0$), all channels remain mutually coherent and the matrix approaches an all-ones form. As the linewidth-delay product increases, the off-diagonal elements decay exponentially, causing the matrix coefficients to localize around its main diagonal. Fig.~\ref{Figure4_2} and \ref{Figure4_3} illustrate this transition for different laser linewidths and differential delays. Increasing either parameter deepens the loss of long-range coherence across the network, indicating that only neighboring OTTD channels remain approximately mutually coherent.

The next subsection presents the Fourier-domain interpretation of laser-induced coherence decay, revealing consequent effects such as coherent RF aperture shrinkage and beam-pattern degradation.

\subsection{Fourier-domain analysis of spatial coherence function reshaping}

By introducing the spatial index variable $x$=$(m-n)$, the mean RF power array factor $\braket{\hat{{P}}_r(\alpha)}$ can be expanded in terms of the spatial coherence function $\tilde{C}_x$ as follows,
\begin{equation}
\braket{\hat{{P}}_r(\alpha)}  = \sum_{x=-(N-1)}^{N-1} \tilde{C}_x e^{jx \gamma_{\alpha}} 
\label{Eq17}
\end{equation}
The spatial coherence function $\tilde{C}_x$ quantifies the average cumulative correlation between all pairs of OTTD channels with delay difference $x \tau$, where the index $x \in [-(N-1), N-1]$. Since an $N$-element array has $(N-|x|)$ number of OTTD channel-pairs separated by the index $x$, having average mutual correlation $\tilde{\rho}^{|x|}$, the coherence function can be expressed as,
\begin{equation}
\tilde{C}_x =  (N-|x|) \tilde{\rho}^{|x|}\ \ \ \ \forall\ \ \ x \in[-(N-1), (N-1)]
\label{Eq18}
\end{equation}
\begin{figure}[ht]
    \centering
    \includegraphics[width=0.8\linewidth]{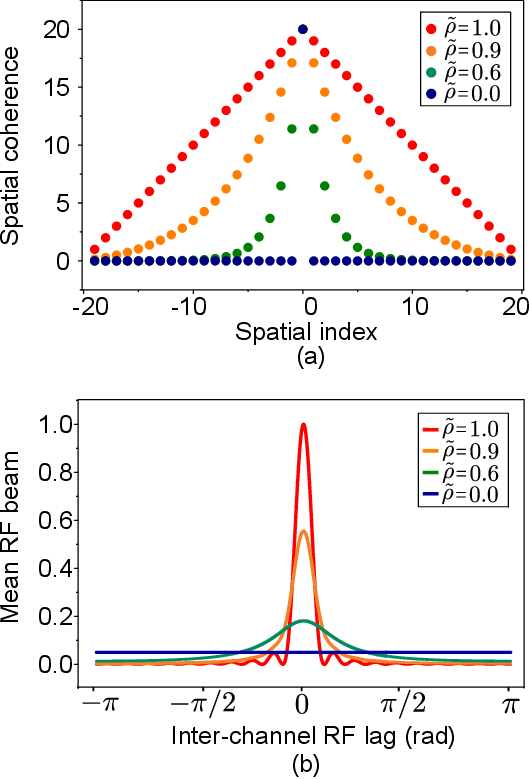}
    \caption{(a) Distribution of the spatial coherence function $\tilde{C}_x$, and (b) the normalized mean RF beam-pattern $\braket{\hat{{P}}_r(\alpha)}/N^{2}$, corresponding to different values of average correlation between neighbouring OTTD channels $\tilde{\rho}$ for a beamforming network of antenna array-dimension $N$=20.}
    \label{FigureFT}
\end{figure}

Equation~(\ref{Eq17}) establishes the centrepiece of this work: the spatial coherence function  and the mean RF beam-pattern form a Fourier-transform pair. Consequently, the spatial coherence distribution across the beamforming network directly determines the resultant beam-pattern. Any reduction in long-range coherence therefore translates into shrinkage of the effective coherent RF aperture, and consequently beam distortion. To verify this argument, we discuss the following scenarios:

\subsubsection{Ideal coherent systems}

In ideal coherent beamforming networks, the constituent OTTD channels are fully correlated to each other i.e. $\tilde{\rho}$=1. For such systems, the spatial coherence function is,
\begin{equation}
        \tilde{C}_x =  (N-|x|)\ \ \ \ \forall\ \ \ x \in[-(N-1), (N-1)]
\label{Eq21}
\end{equation}
From (\ref{Eq21}), it can be interpreted that spatial coherence function has a triangular distribution in the $x$-domain for ideal coherent systems, as reflected in Fig.~\ref{FigureFT}(a) ($\tilde{\rho}$=1, red-dots). Linear slope of the spatial coherence distribution indicates the absence of coherence decay between the OTTD channels of the beamforming network.

The corresponding Fourier-domain representation of $ \tilde{C}_x$ i.e. the distribution of the mean RF power array factor $\braket{\hat{{P}}_r(\alpha)}$ in the $\gamma_{\alpha}$-domain can be derived to be,
\begin{equation}
        \braket{\hat{{P}}_r(\alpha)} = \frac{\sin^{2}{\big(N \gamma_{\alpha}/2 \big)}}{\sin^{2}{\big(\gamma_{\alpha}/2 \big)}}
\label{Eq21_xy}
\end{equation}
The above equation indicates that the mean RF beam-pattern obeys a squared-sinc distribution in the ideal coherent scenario, as reflected by the normalized beam-pattern plot in Fig.~\ref{FigureFT}(b) ($\tilde{\rho}$=1, red-line). It must be pointed out that the RF beam peaks at $\gamma_{\alpha=\alpha_o}$=0 (or integral multiple of $2\pi$ radians), where $\alpha_o$ is the direction of beam maxima. The mean beam peak is $\braket{\hat{{P}}_r(\alpha_o)}$=$N^{2}$, which is essentially the normalization constant used for computing the plots shown in Fig.~\ref{FigureFT}(b).

\subsubsection{Partially incoherent system}

In partially incoherent beamforming networks, the average correlation between neighbouring channels lies in the range $0<\tilde{\rho}<1$. For such systems, the coherence function can be represented as a function of finite linewidth-delay product $\Delta \nu \tau$ as follows,
\begin{equation}
        \tilde{C}_x =  (N-|x|) e^{-\pi \Delta \nu \tau|x|}\ \ \ \ \forall\ \ \ x \in[-(N-1), N-1]
\label{Eq22}
\end{equation}
From the above equation, it can be interpreted that spatial coherence distribution has an exponential profile in partially incoherent systems, as can be verified from Fig.~\ref{FigureFT}(a) ($\tilde{\rho}$=0.9: orange-dots, $\tilde{\rho}$=0.6: green-dots). Exponential tapering of the spatial coherence distribution results in the shrinkage of effective RF aperture of the antenna array. Moreover, the system with poorer inter-channel correlation ($\tilde{\rho}$=0.6) suffers from a greater effective aperture shrinkage compared to the system with relatively larger inter-channel correlation ($\tilde{\rho}$=0.9).

The corresponding Fourier-domain representation of the coherence function for partially incoherent systems can be derived to be,
\begin{equation}
        \braket{\hat{{P}}_r(\alpha)} = N+\Gamma(\tilde{\rho}, \gamma_\alpha)
\label{Eq21_2}
\end{equation}
The average RF beam-pattern characterized by (\ref{Eq21_2}) deviates from ideal squared-sinc distribution (see (\ref{A4}) and (\ref{A4a}) of Appendix-B), as can be verified from the normalized beam-pattern plots in Fig.~\ref{FigureFT}(b) ($\tilde{\rho}$=0.9: orange-lines, $\tilde{\rho}$=0.6: green-lines). It can be observed that the beam-pattern undergoes significant degradation in terms of peak reduction, directivity loss, and mainlobe broadening as a consequence of laser phase noise induced effective RF aperture shrinkage. The system with greater aperture shrinkage ($\tilde{\rho}$=0.6) suffers from greater RF beam degradation.

\subsubsection{Fully incoherent systems}
In completely incoherent beamforming networks, the OTTD channels are uncorrelated to each other i.e.  $\tilde{\rho}$=0. For such systems, the spatial coherence function boils down to,
\begin{equation}
         \tilde{C}_x=  \begin{cases}
N, & \text{for } x=0 \\
0, & \text{for } x \neq 0
\end{cases}
\label{Eq23}
\end{equation}
\begin{figure*}[ht]
    \centering
\includegraphics[width=\linewidth]{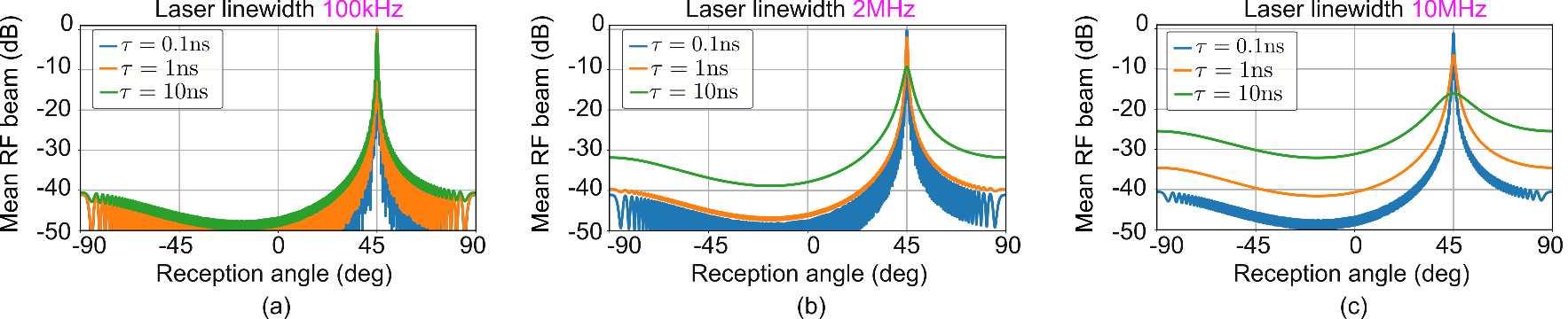}
    \caption{Normalized mean RF beam-pattern $\braket{\hat{{P}}_r(\alpha)}/N^{2}$ (in dB) with maxima oriented along $\alpha_o$=$45^{\circ}$, of an OTTD-based RF beamforming network of array-dimension $N$=256, corresponding to various differential delays $\tau$, driven by a laser of linewidth (a) $\Delta \nu$=100kHz, (b) $\Delta \nu$=2MHz, and (c) $\Delta \nu$=10MHz.}
    \label{Figure7_2}
\end{figure*}
Equation~(\ref{Eq23}) indicates that only self-correlation of the individual OTTD channels in fully incoherent systems contribute to the spatial coherence function. Due to this, the only non-zero component of the spatial coherence function appears for $x$=0, as reflected in Fig.~\ref{FigureFT}(a) ($\tilde{\rho}$=0, blue-dots). This conveys that the effective RF aperture shrinkage is maximum for fully incoherent beamforming networks.

\begin{figure*}[ht]
    \centering
\includegraphics[width=1\linewidth]{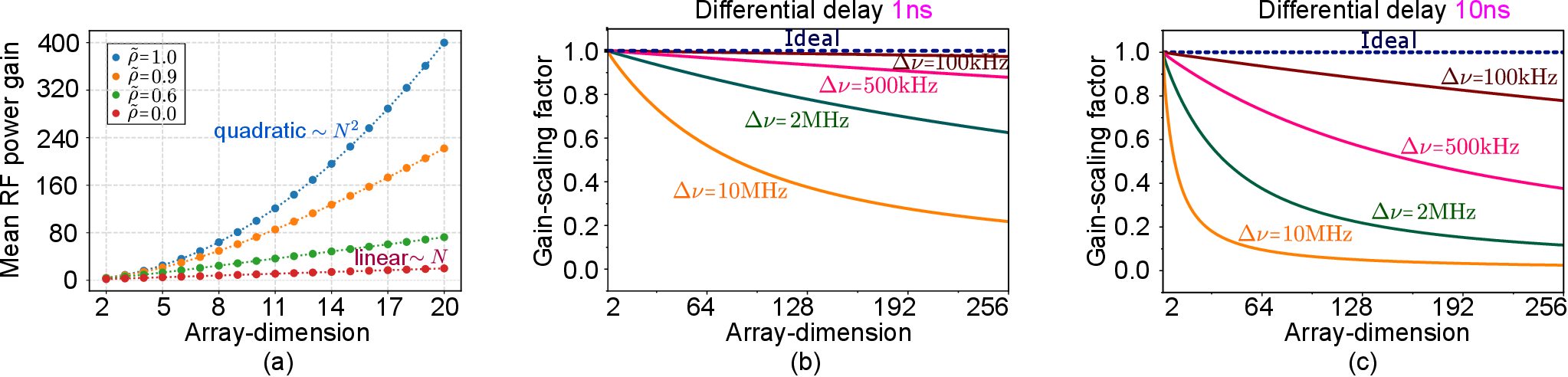}
    \caption{(a) Variation of mean RF power gain $\braket{\hat{G}_r}$ with array-dimension $N$, corresponding to different values of average correlation between neighbouring OTTD channels. Gain-scaling factor $\tilde{g}_r$ versus array-dimension $N$ for different laser linewidths $\Delta \nu$, and differential delay (b) $\tau$=1ns, (c) $\tau$=10ns.} 
    \label{Figure7}
\end{figure*}
\begin{figure*}[ht]
    \centering
\includegraphics[width=\linewidth]{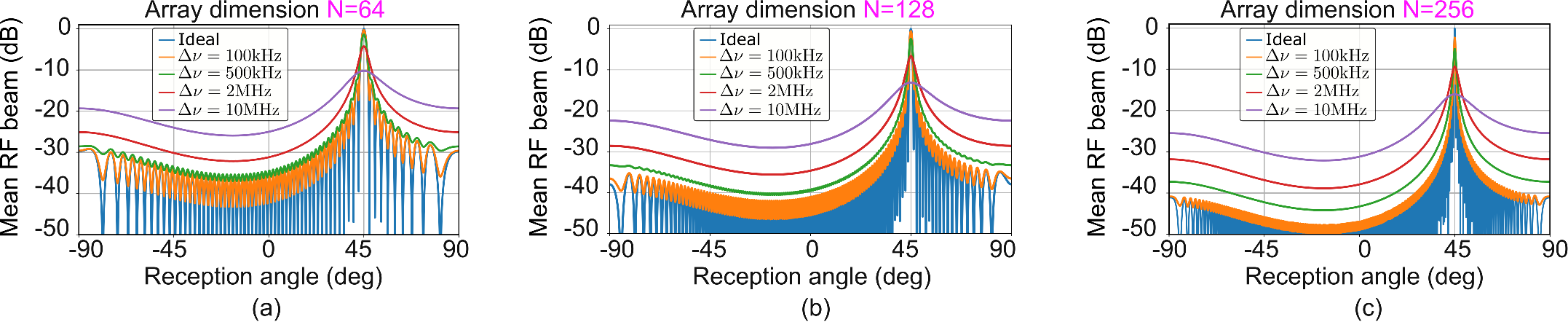}
    \caption{Normalized mean RF beam-pattern with maxima along $\alpha_o$=$45^{\circ}$ for an OTTD-based RF beamforming network with differential delay $\tau$=10ns and array-dimension (a) $N$=64, (b) $N$=128, and (d) $N$=256, corresponding to different laser linewidths $\Delta \nu$.} 
    \label{Figure8}
\end{figure*}

The Fourier-domain representation of the corresponding spatial coherence function can therefore be derived to be,
\begin{equation}
        \braket{\hat{{P}}_r(\alpha)} = N
\label{Eq21_3}
\end{equation}
Equation~(\ref{Eq21_3}) conveys that in fully incoherent systems, the effective RF aperture collapse is so severe that the resultant RF beam-pattern ceases to be a spatially directive function. This can further be verified from the normalized beam-pattern plot in Fig.~\ref{FigureFT}(b) ($\tilde{\rho}$=0, blue-line).

The analysis presented in this subsection demonstrates that the linewidth-delay product governs the shape of the spatial coherence function of a beamforming network, which in turn determines the RF beam-pattern through the Fourier relationship established in (\ref{Eq17}). Increasing either the laser linewidth or the differential delay between neighbouring channels produces the same qualitative effect: progressive aperture shrinkage, lower beam directivity, broader main lobes, and reduced beamforming gain. The beam-patterns represented in Fig.~\ref{Figure7_2} confirm these predictions. Fig.~\ref{Figure7_2}(a) plots the normalized mean RF beam-pattern (in dB) of a partially incoherent system designed to have a beam maxima along $\alpha_o$=$45^\circ$, driven by a laser of linewidth $\Delta \nu$=100kHz for various differential delays. It can be noted that as the differential delay is progressively increased from $\tau$=0.1ns to $\tau$=10ns, beam peak reduction and directivity loss becomes more prominent. These detrimental effects become even more severe when the system is driven by a laser having a relatively broader linewidth of $\Delta \nu$=2MHz, as shown in Fig.~\ref{Figure7_2}(b). A similar trend can be observed for an even broader laser linewidth of of $\Delta \nu$=10MHz in Fig.~\ref{Figure7_2}(c). This illustrates the gradual transition from nearly ideal coherent beamforming to severely degraded operation as the linewidth-delay product is increased. This provides the physical foundation for the performance and scalability analysis presented in the following section.

\section{Beamforming fidelity analysis}

\subsection{RF power gain downscaling}
\begin{table*}[ht]
\centering
\caption{Summary of total mean RF power gain and gain-scaling factor as functions of array-size in different operating regimes}
 \begin{tabular}{||c| c |c | c||} 
 \hline
\textbf{Regime}  & \textbf{Average inter-channel correlation ($\tilde{\rho}$)} & \textbf{Mean RF power gain ($\hat{G_r}$) } & \textbf{Gain-scaling factor ($\tilde{g}_r)$}\\ [0.5ex] 
 \hline
 Ideal coherent & $\tilde{\rho}$=1 & $\braket{\hat{G_r}}$=$N^{2}$ & $\tilde{g}_r$=1 \\ [0.8ex]
  \hline
 Partially incoherent & $0<\tilde{\rho}<1$ & $\braket{\hat{G_r}}$=$N + 2\sum_{x=1}^{N-1} (N-x) \tilde{\rho}^{x}$ & $1/N<\tilde{g}_r<1$\\ [0.8ex]
 \hline
Completely incoherent & $\tilde{\rho}$=0 & $\braket{\hat{G_r}}$=$N$ & $\tilde{g}_r$=$1/N$\\ [0.8ex]
\hline
\end{tabular}
\label{Table1}
\end{table*}

Having studied the impact of laser phase noise on RF beam-pattern in the previous section, we will now quantify its impact on one of the most critical performance metrics of a beamforming network: the achievable RF power gain. For partially incoherent beamforming networks, the average RF power gain  $\braket{\hat{G}_r}$=$\braket{\hat{{P}}_r(\alpha_o)}$=$\braket{|\hat{I}_r(\alpha_o)|^{2}}$ can be derived to be (see Appendix-C),
\begin{equation}
\braket{\hat{G_r}} =  N + 2\sum_{x=1}^{N-1} (N-x) \tilde{\rho}^{x}  
\label{Eq24_2}
\end{equation}
Equation~(\ref{Eq24_2}) indicates that for ideal coherent beamforming networks with zero linewidth-delay product, the average gain quadratically scales up with the array-size i.e. $\braket{\hat{G_r}}$=$N^{2}$. This can be verified from the plot of mean RF power gain versus array-dimension for $\tilde{\rho}$=1 in Fig.~\ref{Figure7}(a). However, (\ref{Eq24_2}) also predicts that for partially incoherent systems with non-zero linewidth-delay product, the average gain deviates from being an quadratic function of array-size. This can be verified for the plot of $\tilde{\rho}$=0.9 and $\tilde{\rho}$=0.6 shown in Fig.~\ref{Figure7}(a). It can also be observed that the aforementioned deviation is greater for the system with $\tilde{\rho}$=0.6, compared to the system with $\tilde{\rho}$=0.9. In the worst-case scenario i.e. for a fully incoherent system with a very large linewidth-delay product, the gain tends to scale linearly with the array-size, as shown in Fig.~\ref{Figure7}(a) for $\tilde{\rho}$=0. Therefore, the linewidth-delay product affects RF gain-scaling of beamforming networks. To highlight this effect, we introduce the parameter called gain-scaling factor, which quantifies the real gain normalized with respect to the ideal gain i.e. $\tilde{g}_r$=$\braket{\hat{G_r}}/N^{2}$. The expression of the gain-scaling factor is,
\begin{equation}
{\tilde{g}_r} =  \frac{1}{N} + \frac{2}{N^{2}}\sum_{x=1}^{N-1} (N-x) \tilde{\rho}^{x} 
\label{Eq24_2new}
\end{equation}

Naturally, the gain-scaling factor for ideal coherent systems is a constant function of the array-size, as reflected in Fig.~\ref{Figure7}(b)-(c). On the contrary, the gain-scaling factor for partially incoherent systems drops as the array-size is increased, as can be verified from Fig.~\ref{Figure7}(b)-(c) as well. Moreover, it can be noted that gain downscaling is even steeper as the laser linewidth and (or) delay is increased. A particularly important result is that gain downscaling asymptotically approaches an inverse dependence on the array-size, $\tilde{g}_r\sim 1/N$, as the system approaches the incoherent regime.

Gain downscaling due to laser phase noise manifests as reduction of normalized mean RF beam peak, as shown in Fig.~\ref{Figure8}(a)-(c). It can be observed that the reduction of normalized beam peak is lower for the beamforming network of array-size $N$=64 in Fig.~\ref{Figure8}(a), compared to when the array-size $N$=128 in Fig.~\ref{Figure8}(b). Following the similar trend, Fig.~\ref{Figure8}(c) shows that beam peak degradation is more severe for a larger array of dimension $N$=256. This analysis shows that by simply increasing the number of antennas in the RF array cannot compensate for coherence loss induced by laser phase noise. Beyond a certain array-size, additional elements contribute progressively less coherent gain while increasing implementation complexity. 

Table~\ref{Table1} summarizes the mean power gain and gain-scaling factor as functions of the array-dimension for different operating regimes categorized based on the average inter-channel correlation. The upcoming subsections investigate the physical reason behind gain downscaling, and the corresponding influence on system scalability and performance.

\subsection{Coherent array gain degradation and fluctuation analysis}

The reduction in total mean gain alone does not fully characterize beamforming fidelity. It is equally important to distinguish between the deterministic array coherent gain and stochastic gain fluctuations introduced by laser phase noise. The total mean RF power gain $\braket{\hat{G}_r}$ discussed in the previous subsection has two constituent components: (i) average coherent array gain ${G}_c$, and (ii) gain variance $\Delta {G}_{v}$. This is represented by the following equation, 
\begin{equation}
\braket{\hat{G}_r}={G}_c + \Delta {G}_{v}
\label{Sec4-1}
\end{equation}
Average coherent array gain $G_c$=$|\braket{\hat{I}_r(\alpha_o)}|^{2}$ is the fraction of the total mean RF gain that is contributed by the coherent interference between RF-modulated optical fields in the OTTD network. The average coherent array gain $G_c$ can be derived to be,
\begin{equation}
G_c= \frac{\tilde{\rho}^{2} +\tilde{\rho}^{2N+2}-2\tilde{\rho}^{N+2}}{1+\tilde{\rho}^{2}-2\tilde{\rho}} 
\label{Sec4-2}
\end{equation}
\begin{figure*}[ht]
    \centering
\includegraphics[width=1\linewidth]{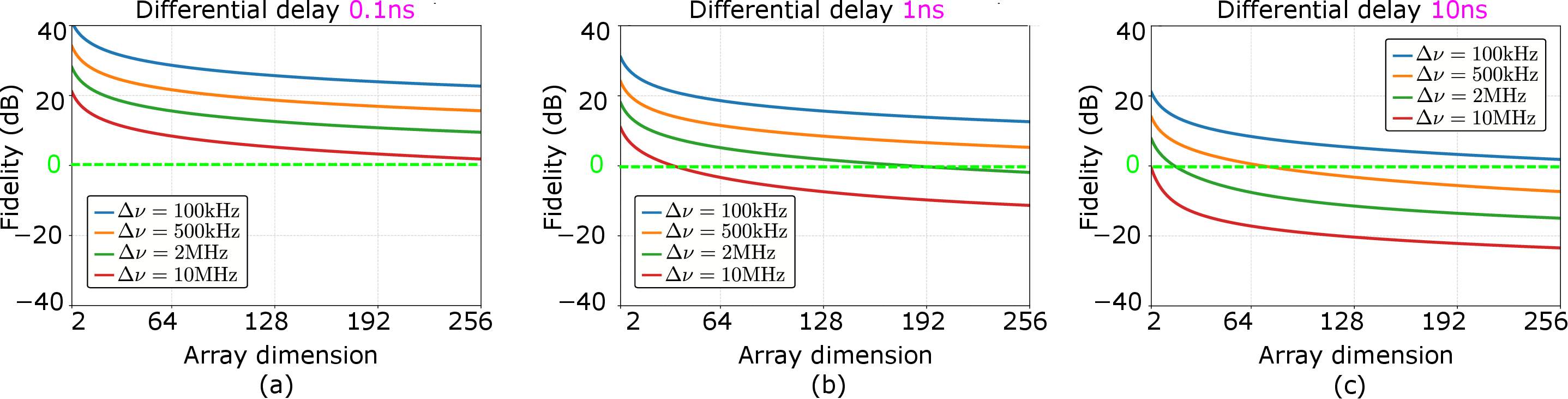}
    \caption{Beamforming fidelity $\tilde{F}$ (in dB) as a function of array-dimension $N$ corresponding to different laser linewidths $\Delta \nu$ when the system has a differential delay of (a) $\tau$=0.1ns, (b) $\tau$=1ns, and (c) $\tau$=10ns.}
    \label{Figure10}
\end{figure*}
On the other hand, gain variance $\Delta G_v$=$\braket{\hat{G}_r}-{G}_c$ is associated with the fraction of total mean RF gain that is dictated by the laser phase noise induced power fluctuations in the OTTD network. Using the expressions of $\braket{\hat{G}_r}$ and $G_c$ from ~(\ref{Eq24_2}) and ~(\ref{Sec4-2}), respectively, the gain variance $ \Delta G_v$ can be derived to be,
\begin{equation}
\Delta G_v = N + 2\sum_{x=1}^{N-1} (N-x) \tilde{\rho}^{x}  -\frac{\tilde{\rho}^{2} +\tilde{\rho}^{2N+2}-2\tilde{\rho}^{N+2}}{1+\tilde{\rho}^{2}-2\tilde{\rho}} 
\label{Sec4-6}
\end{equation}

We quantify beamforming fidelity by introducing the parameter $\tilde{F}$, which is essentially the ratio of average coherent array gain and gain variance,
\begin{equation}
\tilde{F}=\frac{G_c}{\Delta G_v}
\label{Sec4-6new}
\end{equation}
Fig.~\ref{Figure10}(a)-(c) show that the fidelity decreases with increasing linewidth, differential delay, and array-size. The green dotted-lines correspond to the threshold limit at which $\tilde{F}$=0$\mathrm{dB}$ when $G_c$=$\Delta{{G}}_{v}$. Once the fidelity crosses the 0dB threshold, coherent beamforming is no longer possible, as the stochastic fluctuations dominate over the useful coherent array gain.

It is even more interesting to observe from Fig.~\ref{Figure10}(a)-(c) that $\tilde{F}$ approaches the 0$\mathrm{dB}$-limit as the array-size is increased. This happens because in incoherent systems of higher array-dimension, there are relatively larger number of  uncorrelated OTTD channels. As a result, increasing the array-dimension causes coherent array gain to drop and gain variance to increase, which in turn causes the fidelity to drop and eventually hit the 0$\mathrm{dB}$-limit. 

Sections~IV-A and IV-B demonstrate that laser phase noise fundamentally limits the performance of OTTD-based RF beamforming networks through three closely related mechanisms: reduced coherent array gain, increased gain fluctuations, and degraded beamforming fidelity. Together, these effects impose practical limits on the allowable linewidth-delay product and the maximum scalable array-dimension. Building on these results, the next subsection consolidates the analysis into practical performance bounds and design guidelines that relate laser linewidth, differential delay, and array-size.

\subsection{Performance bounds and scalability limits}
The preceding analysis demonstrates that laser linewidth, differential delay, and array-dimension jointly determine beamforming performance. We now combine these parameters into practical design maps that identify the operating regimes and scalability limits of OTTD beamforming networks.

\begin{figure}[ht]
    \centering
\includegraphics[width=0.8\linewidth]{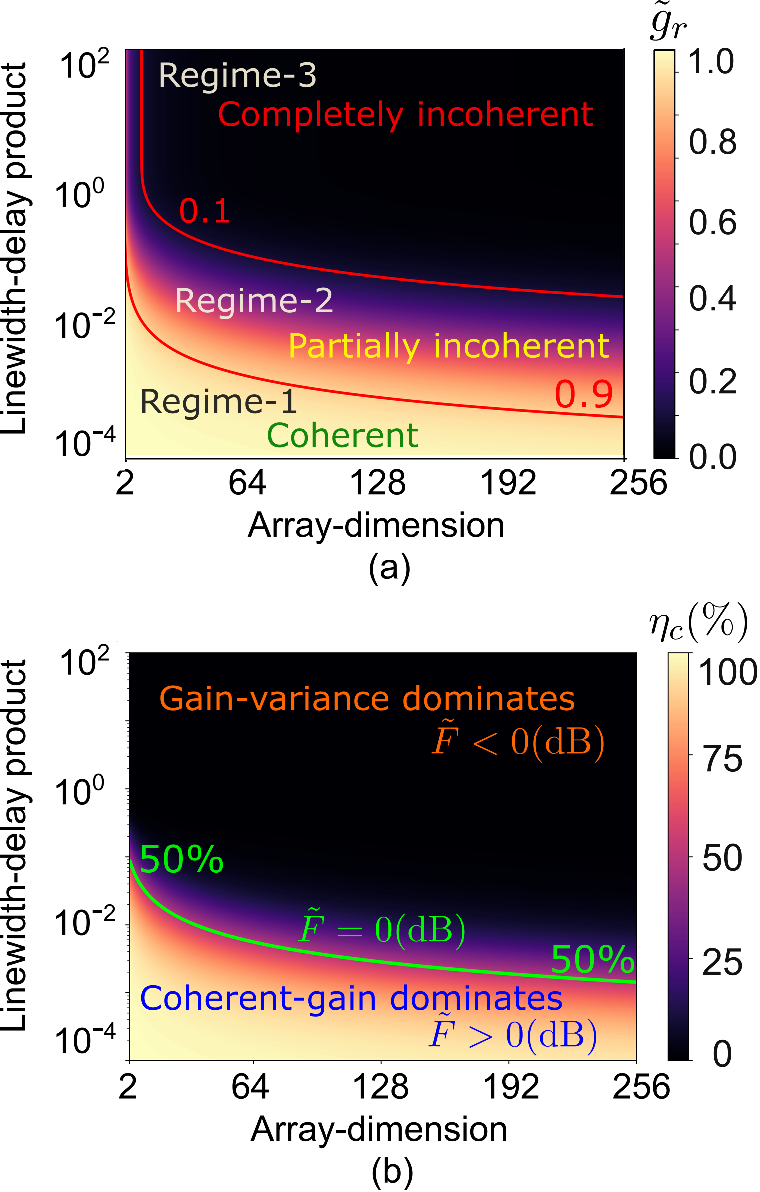}
    \caption{Contour plot of (a) gain-scaling factor $\tilde{g}_r$, and (b) coherent-gain ratio $\eta_{c}$ ($\%$) as a joint function of linewidth-delay product $\Delta \nu \tau$ and array-dimension $N$. }
    \label{Figure10_contour}
\end{figure}

Fig.~\ref{Figure10_contour}(a) summarizes the normalized total average RF power gain as a joint function of linewidth-delay product and array-size. Using a 10$\%$ deviation from the ideal gain as a design criterion, the parameter space naturally separates into (i) coherent ($\tilde{g}_r \geq 0.9$), (ii) partially coherent ($0.1<\tilde{g}_r<0.9$), and (iii) incoherent  ($\tilde{g}_r \leq 0.1$) operating regimes.

Fig.~\ref{Figure10_contour}(b) complements this analysis by showing coherent-gain ratio $\eta_c$=$(G_c/\braket{\hat{G}_r})$ as a joint function of linewidth-delay product and array-size. From (\ref{Sec4-1}) and (\ref{Sec4-6new}), it can be verified that the 0dB limit of coherence-to-variance ratio is met when the coherent-gain is $50\%$ of the gain variance. This in turn indicates that
(i) average coherent array gain dominates over gain variance when 
$\tilde{F}>$0$\mathrm{dB}$ as $\eta_c>$$50\%$, and  (ii) vice-versa when $\tilde{F}<$0$\mathrm{dB}$ as $\eta_c<$$50\%$. The presented contour map in Fig.~\ref{Figure10_contour}(b) identifies the combinations of linewidth-delay product and array-size for which coherent beamforming remains dominant over stochastic gain fluctuations. 

These contour maps provide straightforward design guidelines. For a given laser linewidth, they determine the maximum differential delay, and therefore the maximum beam steering range that can be supported while maintaining a desired beamforming fidelity. Conversely, for a prescribed delay architecture, the same maps specify the minimum laser coherence required to realize a target array-dimension. They therefore serve as practical design tools for future large-scale photonic beamforming systems.

\section{Conclusion}

This paper developed a stochastic framework for investigating the impact of finite laser linewidth on OTTD-based RF beamforming networks. Unlike conventional deterministic analysis, the proposed formulation modeled the RF array factor as a stochastic process, revealing that laser phase noise fundamentally reshapes the spatial coherence function of photonic beamforming networks. Through a Fourier-domain analysis, it was shown that the reshaping of spatial coherence function causes effective aperture reduction, degraded beam directivity, reduced coherent array gain, and increased gain fluctuations. This is the core novelty of this work. The analysis further demonstrated that these effects become increasingly pronounced as the differential delay and array-dimension increase, ultimately limiting the scalability of photonic beamforming networks. By quantifying the interplay among laser linewidth, differential delay, and array-size, this work establishes fundamental performance limits and provides a rigorous theoretical basis for the design of large-scale photonic beamforming systems for next-generation wireless communications, sensing, radar, and radio astronomy. The developed framework provides a foundation for incorporating laser phase noise into the design and optimization of future large-scale photonic beamforming networks, where coherence, rather than bandwidth alone, becomes a fundamental determinant of system scalability.

\appendices

\section{Derivation of the mean RF Power Array Factor}
This appendix derives the closed-form expression for the mean RF power array factor $\braket{\hat{{P}}_r(\alpha)}$ discussed in Section~III-A. We first compute the stochastic average of the random process  expressed in (\ref{Eq8}) as follows,
\begin{equation}
\braket{\hat{{P}}_r(\alpha)} = \sum_{m=1}^{N}  \sum_{n=1}^{N}   \braket{e^{j (\hat{\phi}(t-\tau_m) - \hat{\phi}(t-\tau_n))}} e^{j (m-n) \gamma_{\alpha} } 
\label{A1}
\end{equation}
In the above equation, the exponential term ${e^{j (\hat{\phi}(t-\tau_m)-\hat{\phi}(t-\tau_n))}}$ is the random process associated with the optical phase decorrelation between the $m^{\mathrm{th}}$ and $n^{\mathrm{th}}$ OTTD channels of the beamforming network. In that, the quantity $\hat{\phi}(t-\tau_m) - \hat{\phi}(t-\tau_n)$ is a Wiener random process having zero mean and a variance of $\sigma^{2}_\phi$=$2\pi \Delta \nu |m-n|\tau$ \cite{okoshi1988coherent}. Applying the following Gaussian distribution identity \cite{ishimura2021impact}: $\braket{e^{j (\hat{\phi}(t-\tau_m) - \hat{\phi}(t-\tau_n))}}$=$e^{-\sigma^{2}_\phi/2}$  in ~(\ref{A1}), the closed-form expression of the mean RF power array factor $\braket{\hat{{P}}_r(\alpha)}$ can be derived to be,   
\begin{equation}
\braket{\hat{{P}}_r(\alpha)} = \sum_{m=1}^{N}  \sum_{n=1}^{N}   \tilde{\rho}_{mn}  e^{j (m-n) \gamma_{\alpha} }
\label{A2}
\end{equation}
where the parameter $\tilde{\rho}_{mn}$ in the above equation is equal to $e^{-\pi \Delta \nu |m-n| \tau}$. This result shows that the mean RF power array factor is completely determined by the average correlation between the OTTD channels present in the network.

\section{Derivation of the Mean RF beam-pattern for partially incoherent systems}
This appendix derives the closed-form expression of the mean RF power array factor i.e. mean beam-pattern for partially incoherent systems. Substituting the expression of $\tilde{C}_x$ from (\ref{Eq18}) into (\ref{Eq17}), the resultant expression of $\braket{\hat{{P}}_r(\alpha)}$ can be aligned in the following manner,
\begin{equation}
\begin{split}
\braket{\hat{{P}}_r(\alpha)} 
& = N+2\Re \big \{ \sum_{x=1}^{N-1} (N-x) \tilde{\rho}^{x} e^{jx\gamma_\alpha} \big \}\\
& = N+\Gamma(\tilde{\rho}, \gamma_\alpha)
\end{split}
\label{A4}
\end{equation}
The closed-form expression of the auxiliary function $\Gamma(\tilde{\rho}, \gamma_\alpha)$ can be derived by by evaluating the following finite geometric series: 
\begin{equation}
\begin{split}
\Gamma(\tilde{\rho}, \gamma_\alpha) & =  2\Re \bigg \{N\sum_{x=1}^{N-1} (\tilde{\rho} e^{j\gamma_\alpha})^{x} - \sum_{x=1}^{N-1} x(\tilde{\rho} e^{j\gamma_\alpha})^{x} \bigg \} \\
  & = 2\Re\bigg \{ \frac{(N-1)\tilde{\rho} e^{j\gamma_\alpha} -N (\tilde{\rho} e^{j\gamma_\alpha})^{2} + (\tilde{\rho} e^{j\gamma_\alpha})^{N+1}}{(1-\tilde{\rho} e^{j\gamma_\alpha})^{2}} \bigg \}
\end{split}
\label{A4a}
\end{equation}
Plugging (\ref{A4a}) into (\ref{A4}) establishes the Fourier-domain relationship between the spatial coherence distribution and the mean RF beam-pattern for partially incoherent systems discussed in Section~III-B.

\section{Derivation of the Average RF Power Gain}
This appendix derives the closed-form expression for the mean RF power gain $\braket{\hat{{G}}_r}$. This can be obtained by evaluating the mean RF power array factor in (\ref{A4}) at the beam maxima $\alpha_o$ as follows, 
\begin{equation}
\braket{\hat{{G}}_r}  = N+\Gamma(\tilde{\rho}, \gamma_{\alpha_{o}})
\label{A5_0}
\end{equation}

The expression of the auxiliary function $\Gamma(\tilde{\rho}, \gamma_{\alpha_{o}})$ at the beam maxima can be derived by substituting $\gamma_{\alpha_o}=2k\pi$ ($k$ is an integer) into (\ref{A4a}) to get,
\begin{equation}
\begin{split}
\Gamma(\tilde{\rho}, \gamma_{\alpha_{o}})  & =2 \bigg \{N\sum_{x=1}^{N-1} \tilde{\rho} ^{x} - \sum_{x=1}^{N-1} x\tilde{\rho} ^{x} \bigg \}\\
& =2 \sum_{x=1}^{N-1} (N-x) \tilde{\rho}^{x} 
\end{split}
\label{A5_01}
\end{equation}
Substituting (\ref{A5_01}) in (\ref{A5_0}), we get the following expression,
\begin{equation}
\braket{\hat{{G}}_r}  = N+2 \sum_{x=1}^{N-1} (N-x) \tilde{\rho}^{x}
\label{A5}
\end{equation}
The above equation gives the closed-form expression of mean RF power gain $\braket{\hat{{G}}_r}$ presented in (\ref{Eq24_2}) of Section~IV-A.

\section{Derivation of average coherent array gain}
In this appendix section, we will derive the closed-form expression for average coherent array gain $G_c$=$|\braket{\hat{I}_r(\alpha_o)}|^{2}$. To proceed, we first derive the mean RF field array factor $\braket{\hat{{I}}_r(\alpha_o)}$ along the beam maxima $\alpha_o$. We substitute $\gamma_{\alpha_o}=2k\pi$ ($k$ is an integer) into (\ref{Eq7_1}), and apply the same Gaussian distribution identity (assuming $\tau_c$=0) previously used in Appendix-A, to get,
\begin{equation}
\braket{\hat{{I}}_r(\alpha_o)} = \sum_{n=1}^{N}  \tilde{\rho}^{n}
\label{AppD_2}
\end{equation}
Using the derived expression of $\braket{\hat{{I}}_r(\alpha_o)}$ in (\ref{AppD_2}), the expression of $G_c$=$|\braket{\hat{I}_r(\alpha_o)}|^{2}$ can be derived using the properties of geometric series as follows,
\begin{equation}
\begin{split}
G_c  & = |\sum_{n=1}^{N}  \tilde{\rho}^{n}\ |^{2}\\
& = \frac{\tilde{\rho}^{2} +\tilde{\rho}^{2N+2}-2\tilde{\rho}^{N+2}}{1+\tilde{\rho}^{2}-2\tilde{\rho}} 
\end{split}
\label{AppD_3}
\end{equation}
The above equation gives the closed-form expression of average coherent array gain $G_c$ presented in (\ref{Sec4-2}) of Section IV-B.

%\begin{equation}
   % \theta_i (t) = a_i \delta \theta |E_\mathrm{m}| \cos{(\omega_{\mathrm{m}}t+b_i+r)}
%\label{Eq9}
%\end{equation}

%\bibliography{apssamp}
%\bibliographystyle{ieeetr}
%\bibliography{Bibliography}
%\bibliographystyle{style=numeric-comp}

%apsrev4-2.bst 2019-01-14 (MD) hand-edited version of apsrev4-1.bst
%Control: key (0)
%Control: author (8) initials jnrlst
%Control: editor formatted (1) identically to author
%Control: production of article title (0) allowed
%Control: page (0) single
%Control: year (1) truncated
%Control: production of eprint (0) enabled
\providecommand{\noopsort}[1]{}\providecommand{\singleletter}[1]{#1}%

\end{document}